\documentclass{article}
\usepackage{spconf,amsmath,graphicx,booktabs,amssymb}
\usepackage[hidelinks]{hyperref}

\newcommand{\wer}{WER}
\newcommand{\ours}{SoniGAN}

\title{SYNTHESIZING ULTRASONIC ECHO PROFILES FROM TEXT FOR\\
       OPEN-VOCABULARY SILENT SPEECH RECOGNITION}

\name{Jiacheng Liu, Ruidong Zhang, François Guimbretière, Cheng Zhang}
\address{Cornell University, Ithaca, NY, USA}

\begin{document}
\ninept
\maketitle

\begin{abstract}
Eyewear that senses articulation with inaudible ultrasound can now recognize silent speech over an open vocabulary, but its accuracy is limited by the cost of training data. We present \ours{}, which synthesizes the ultrasonic echo profiles of silent articulation from text and uses them as synthetic supervision. A voice-cloned text-to-speech (TTS) model reads a text corpus, and a two-stage generator synthesizes echo profiles from representations of the spoken content of that audio. Pretraining a recognizer on 104 hours of the synthesized data improves the word error rate (\wer{}) of a public open-vocabulary benchmark by 2.3 points. Adding the sessions the corpus discards for a faulty microphone brings the system to 22.2\% without any language model. Lexicon-constrained decoding, LLM rescoring, and ensembling further bring it to 14.9\%, the first minimally obtrusive silent speech interface to fall inside the error range in which users report no loss of willingness to use silent speech input.

\end{abstract}

\section{INTRODUCTION}
\label{sec:intro}

Silent speech interfaces (SSIs) recover what a person is articulating without any audible sound, serving both those who cannot phonate and anyone who needs speech input where speaking aloud is impractical or intrusive~\cite{pandey2021acceptability}. So far, accuracy has come at the cost of obtrusiveness. Open vocabularies have been reached with facial electrodes~\cite{gaddy2020digital, monalisa} and with ultrasound tongue imaging~\cite{ribeiro2021silent}, both of which put hardware on the skin or hold a probe against the jaw, whereas minimally obtrusive sensing on an ordinary eyeglass frame has stayed with closed command sets~\cite{echospeech}. SoniSpeech~\cite{sonispeech} lifted that restriction, recognizing open-vocabulary silent speech from echo profiles captured with inaudible FMCW ultrasound on eyewear, but at 26.3\% \wer{} it sits well above the range in which users' willingness to adopt silent speech input holds at the level of an error-free system~\cite{pandey2021acceptability}.

Lowering that error rate calls for more paired data, which is exactly what this modality lacks. Every sentence in an echo-profile corpus has to be articulated silently by a wearer with the device on, which makes collection slow and hard to scale. Unpaired text carries no such cost, and generating training data from it has already been exploited for lip reading~\cite{duallip, synthvsr, lipgen} and for EMG-based speech recognition~\cite{chen2025confidence, stegan}. However, no generative model exists for ultrasonic echo profiles, and whether synthesized profiles can improve recognition performance is untested.

To close the gaps, we present \ours{}. It generates the echo profiles of silent articulation from text, providing synthetic supervision for silent speech recognition. The system leverages a voice-cloned TTS model that reads unpaired text; the audio is reduced to soft speech units~\cite{softunits} that keep what is said and discard who says it, and a generator turns those units into echo profiles on which the recognizer is pretrained. The articulatory cue sits in both the coarse envelope of an echo profile and its fringe structure. The generator therefore produces the two in sequence, the envelope first and then the fringes. Each stage is supervised by a frozen recognizer and by two critics that judge local texture and temporal structure separately. 

We make two contributions. \textbf{(1)} \textbf{The first text-to-echo-profile generator, whose output serves as synthetic supervision for silent speech recognition.} A judge recognizer trained on real data reads the generated profiles at 29.0\% \wer{}, against 30.4\% for the real profiles of the same sentences. Pretraining on 104\,h of synthesized data and fine-tuning on real profiles then lowers \wer{} from 25.9\% to 23.6\%. \textbf{(2)} \textbf{The first echo-profile system below 15\% \wer{}.} Recovering the sessions SoniSpeech discards for a faulty microphone~\cite{sonispeech} brings the system to 22.2\% without a language model, and lexicon-constrained decoding, LLM rescoring and ensembling bring it to 14.9\%, to our knowledge the first time a minimally obtrusive SSI falls inside the error range in which users report no loss of willingness to use silent speech input~\cite{pandey2021acceptability}.

% This yields two results. \textbf{Echo profiles can be synthesised from text, and the result supervises rather than merely resembles real data.} A recognizer trained on real profiles reads the generated ones at 29.0\% \wer{} against 30.4\% for the recordings they imitate, and pretraining on 104\,h of them, followed by fine-tuning on real recordings, lowers \wer{} by 2.3 points to 23.6\%, where noise and replayed recordings matched for optimiser steps gain nothing. 

% \textbf{A minimally obtrusive SSI falls below 15\% \wer{}.} Utilizing the sessions SoniSpeech discards for a faulty microphone~\cite{sonispeech}, and adding contrastive learning, lexicon-constrained decoding, LLM rescoring and ensembling, brings the system to 14.9\%. To our knowledge, this is the first time the \wer{} of a minimally obtrusive SSI falls inside the error range in which users report no loss of willingness to use silent speech input~\cite{pandey2021acceptability}.
\section{RELATED WORK}
\label{sec:prior}

\begin{figure*}[t]
  \centering
  \includegraphics[width=\textwidth]{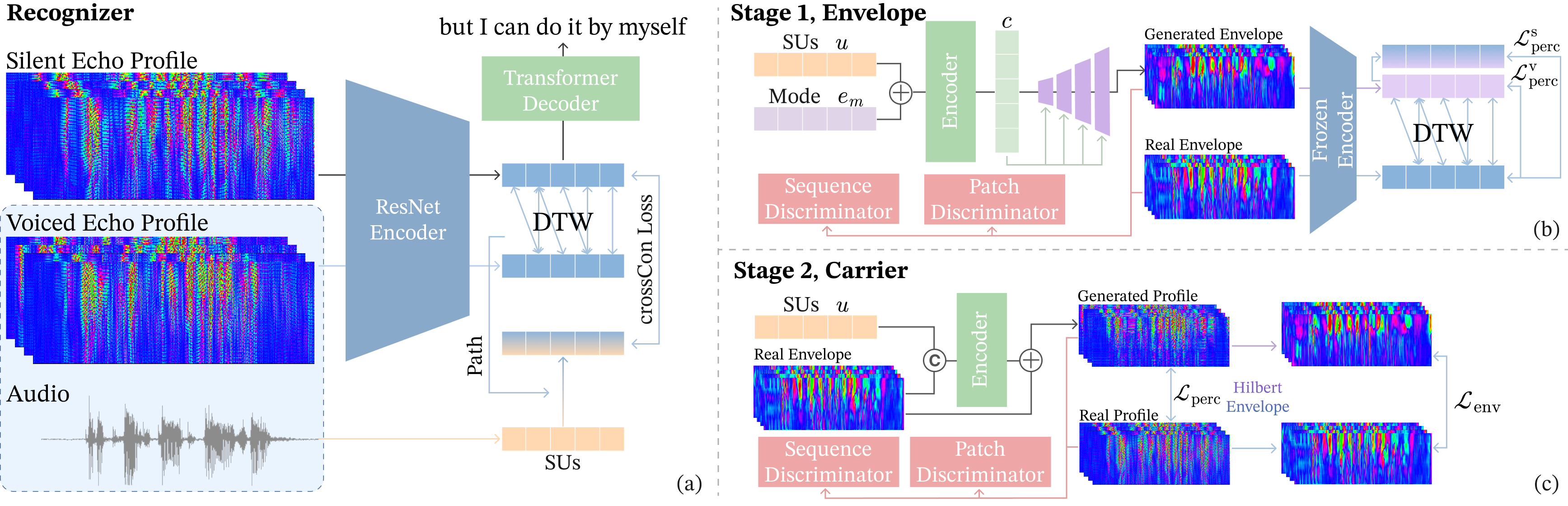}
  \caption{Overview of \ours{}. (a) The recognizer encodes echo profiles with a ResNet encoder and decodes text with a Transformer; crossCon contrasts its output against soft speech units $u$, with a DTW path aligning the silent profile to its voiced counterpart. (b) Stage~1 encodes $u$ and the mode vector $e_m$ into $c$ and expands it into an envelope, under two discriminators and $\mathcal{L}_{\text{perc}}$ from a frozen encoder. (c) Stage~2 concatenates $u$ with an envelope and predicts the residual that carries the fringe structure, under $\mathcal{L}_{\text{perc}}$, $\mathcal{L}_{\text{env}}$ on the Hilbert envelope, and a second pair of discriminators. The circled plus denotes element-wise addition and the circled C concatenation along the channel axis.}

  \label{fig:system}
\end{figure*}

\subsection{Open-vocabulary Silent Speech Recognition on Wearables}
\label{ssec:prior_ssi}

Gaddy and Klein~\cite{gaddy2020digital, gaddy2021improved} pioneered open-vocabulary silent speech recognition with an approximately 20-hour, 8-channel facial EMG corpus, which has since become the standard benchmark for the modality. Benster \textit{et al.}~\cite{monalisa} align silent EMG with recorded voiced sessions through a cross-modal contrastive objective and rescore with an LLM, reducing word error rate to 12.2\%. Ultrasound tongue imaging reaches continuous speech from a different angle. Ribeiro \textit{et al.}~\cite{ribeiro2021silent} recognise multi-speaker silent speech from synchronised tongue imaging and lip video, and report a substantial domain gap between silent and modal articulation that adaptation only partly closes. Both modalities, however, require skin-contact electrodes or a head-stabilised probe. Active acoustic sensing removes that burden by probing articulation with inaudible sweeps from an ordinary eyeglass frame~\cite{echospeech}, at the cost of a closed command set, until SoniSpeech~\cite{sonispeech} paired the same front-end with an open-vocabulary corpus and a CTC recognizer, whose baseline remains at 26.3\% \wer{}.

\subsection{Learning from Synthetic Supervision in SSI}
\label{ssec:prior_synth}

Synthetic supervision has emerged as a common remedy for the data scarcity of silent speech interfaces. In lip reading, DualLip~\cite{duallip} generates video--text pairs to exploit low-resource corpora, SynthVSR~\cite{synthvsr} scales this idea to 3{,}652 hours of lip videos rendered from speech audio, and LipGen~\cite{lipgen} employs speech-driven visual synthesis to improve robustness in real-world settings. Beyond the visual modality, Chen \textit{et al.}~\cite{chen2025confidence} train an EMG-to-speech model on signals produced by the speech-conditioned generator STE-GAN~\cite{stegan}.

Compared to video and EMG synthesis, \ours{} addresses a distinct condition. Real data for this modality is far scarcer, and range compression places the articulatory cue in fringe structure that a perceptually plausible sample does not reproduce, which, to our knowledge, no prior work has attempted to synthesize.
\section{METHODS}
\label{sec:methods}

\subsection{Recognizer}
\label{ssec:recog}

As shown in Figure~\ref{fig:system}\,(a), the recognizer is a ResNet-18 followed by a six-layer Transformer encoder~\cite{vaswani2017attention} ($d_{\text{model}}{=}384$, 8 heads, relative positional attention), 22.4\,M parameters in total. To utilize the audible speech provided in the SoniSpeech dataset~\cite{sonispeech}, which the baseline leaves unused, we exploit it with the cross-contrast (crossCon) loss of Benster \textit{et al.}~\cite{monalisa}, which ties an echo-profile encoder to a speech representation frame by frame. Rather than training a separate audio branch as in the original work, we contrast the ResNet-18 output directly against soft speech units pooled to the same 12.5\,Hz rate, so no extra parameters are introduced.

CrossCon is an InfoNCE objective in which simultaneous frames of the two representations form the positive pair and every other frame of either modality in the batch is a negative. A silent recording has no audible speech of its own, so its units must come from the voiced recording of the same sentence. The two takes differ in timing, so we align them with a DTW path computed between their echo-profile representations and warp the speech units onto the silent frames along that path. The recognizer minimizes $\mathcal{L}_{\text{CTC}} + \mathcal{L}_{\text{con}}$.

\subsubsection{Condition Corpus Preparation}

Existing speech corpora~\cite{panayotov2015librispeech} differ from SoniSpeech in content and literary style, since SoniSpeech is built on the social-dialogue corpus SODA~\cite{kim2023soda}, and conditioning on them would introduce out-of-vocabulary (OOV) words. We therefore draw text from two movie-dialogue and subtitle datasets~\cite{danescu2011chameleons, lison2016opensubtitles2016}, keeping sentences of 10--25 words whose vocabulary is covered by the training set and discarding any sentence that appears in the SoniSpeech evaluation transcripts, which yields 76{,}382 sentences.

A flow-matching TTS model, F5-TTS~\cite{chen2025f5}, is cloned on 11.7 hours of the corpus speaker's own voiced recordings and reads this text, producing 104 hours of audio. Its speed parameter is drawn per utterance from a calibrated distribution so that the synthetic corpus matches the speaking rate of the real recordings; the two measure $2.69 \pm 0.40$ and $2.71 \pm 0.37$ words/s.

Following STE-GAN~\cite{stegan}, the generator is conditioned not on this audio but on its soft speech units~\cite{softunits}, a 256-d vector every 20\,ms (50\,Hz) that retains what is being said and discards who is saying it. This further removes the timbre difference between vocoder-generated and real audio. We write $u$ for the unit sequence of an utterance.

\subsubsection{Echo Profile Synthesis}

The SoniSpeech device emits inaudible FMCW sweeps and range-compresses the reflections into \emph{echo profiles}, four channels of 80 range bins at 200\,Hz that map how the reflecting surfaces of the face move over time~\cite{sonispeech}. The articulatory cue sits in both the coarse envelope of an echo profile and the fine fringe structure left by the carrier, which a generator is unlikely to recover in a single step. We therefore synthesize the two in sequence with two GANs, predicting the envelope first and the carrier conditioned on it (Fig.~\ref{fig:system}a).

SoniSpeech provides echo profiles in two modes, silent and voiced, the former recorded while the participant articulates without phonation and the latter while speaking the utterance aloud. To exploit both while still directing the generator towards a target mode, each mode $m\in\{\mathrm{v},\mathrm{s}\}$ is represented by a learnable vector $e_m\in\mathbb{R}^{256}$ that conditions the generator. We write $a_m$ for a real envelope and $y_m$ for the full echo profile it comes from.

\textbf{Stage 1, envelope.} The generator produces $\hat{a}_m = G_1(u, e_m)$, a $4\times80\times T$ envelope at 200\,Hz, from the units $u$ of the utterance's voiced recording (Fig.~\ref{fig:system}b). It is trained on 256-frame windows (1.28\,s, 64 unit frames) and applied to whole sentences at synthesis. The mode vector $e_m$ is broadcast over time and added to $u$, and an encoder of four residual dilated 1-D convolutions turns the sum into a latent sequence $c$; a decoder then grows a 5-bin seed computed from $c$ to 80 range bins at 200\,Hz, every block modulated by $c$, following the StyleGAN2-style decoder of SynthVSR~\cite{synthvsr}.

Each mode is assigned two critics $D^{k}_{m}$, $k\in\{\mathrm{p},\mathrm{q}\}$, one for the realism of local detail and one for temporal structure. The patch critic $D^{\mathrm{p}}_{m}$~\cite{isola2017image} scores patches of 34 range bins $\times$ 34 frames and judges local texture and amplitude statistics; since plausible texture is plausible wherever it sits, the sequence critic $D^{\mathrm{q}}_{m}$ pools over range and judges temporal structure instead. Both compare $\hat{a}_m$ with real envelopes of the same mode under the least-squares objective~\cite{mao2017least}, giving the generator the term
\begin{equation}
\label{eq:ladv}
\mathcal{L}_{\text{adv}} = \frac{1}{8} \sum_{m} \sum_{k} \mathbb{E}\bigl[(D^{k}_{m}(\hat{a}_m) - 1)^{2}\bigr],
\end{equation}
the standard least-squares generator loss averaged over the four critics.

Matching the style of a given mode, however, does not guarantee that the generated envelope carries the intended content. We therefore adopt the perceptual loss of SynthVSR~\cite{synthvsr}. A recognizer is first trained on real envelopes following Section~\ref{ssec:recog} and then frozen, supervising content through its ResNet-18 output $z(\cdot)$ and its CTC posteriors $p(\cdot)$,

\begin{equation}
\label{eq:lperc}
\mathcal{L}^{m}_{\text{perc}} = \lVert z(\hat{a}_m) - z(a_m) \rVert_{1} + \lambda_{\text{logit}}\, \mathrm{KL}\bigl(p(a_m) \,\Vert\, p(\hat{a}_m)\bigr),
\end{equation}

with $\lambda_{\text{logit}} = 0.1$. Both terms compare frames pairwise, so $\hat{a}_m$ and $a_m$ must share a clock. This holds for the voiced mode, whose units and envelope come from one recording, but a silent envelope $\hat{a}_{\mathrm{s}}$ is generated from units of the voiced take of the same sentence. We align the two as in Section~\ref{ssec:recog}, warping $a_{\mathrm{s}}$ onto $\hat{a}_{\mathrm{s}}$ along a DTW path computed between the two real envelopes in the latent space of the frozen recognizer. The generator minimizes
\begin{equation}
\label{eq:lg}
\mathcal{L}_{G_1} = \lambda_{\mathrm{v}} \mathcal{L}^{\mathrm{v}}_{\text{perc}} + \lambda_{\mathrm{s}} \mathcal{L}^{\mathrm{s}}_{\text{perc}} + \lambda_{\text{adv}} \mathcal{L}_{\text{adv}}.
\end{equation}

\textbf{Stage 2, carrier.} As shown in Fig.~\ref{fig:system}c, the second generator $G_2$ predicts the residual between an envelope and the profile it came from, that is, the fringe structure the envelope lacks,
\begin{equation}
\label{eq:g2}
\hat{y}_m = a_m + G_2(a_m, u).
\end{equation}
The units $u$ are embedded by a $1{\times}1$ convolution, upsampled to 200\,Hz and broadcast over range, then concatenated with $a_m$ and passed through a stack of dilated 2-D residual convolutions with no downsampling, since the fringes are a high spatial frequency along range that striding would erase. An $L_1$ loss on the range-axis Hilbert envelope keeps the envelope of $\hat{y}_m$ faithful to that of $y_m$, while the fringe structure is restored by a perceptual loss through a second frozen recognizer, trained on intact echo profiles with latent $z'(\cdot)$, and by a patch and a sequence critic of the Stage~1 design,

% \begin{equation}
% \label{eq:lg2}
% \mathcal{L}_{G_2} = \lambda_{\text{env}} \mathcal{L}_{\text{env}} + \lambda'_{\text{perc}} \mathcal{L}'_{\text{perc}} + \lambda'_{\text{adv}} \mathcal{L}'_{\text{adv}},
% \end{equation}

% where $\mathcal{L}_{\text{env}} = \bigl\lVert |\mathcal{H}(\hat{y}_m)| - |\mathcal{H}(y_m)| \bigr\rVert_{1}$ and $\mathcal{L}'_{\text{perc}}$ is the latent term of Eq.~\eqref{eq:lperc} taken through $z'$.

\begin{equation}
\label{eq:lg2}
\mathcal{L}_{G_2} = \lambda_{\text{env}} \mathcal{L}_{\text{env}} + \lambda'_{\text{perc}} \mathcal{L}_{\text{perc}} + \lambda'_{\text{adv}} \mathcal{L}_{\text{adv}},
\end{equation}

where $\mathcal{L}_{\text{env}} = \bigl\lVert |\mathcal{H}(\hat{y}_m)| - |\mathcal{H}(y_m)| \bigr\rVert_{1}$.

\subsection{Decoding}
\label{ssec:decoding}

Greedy CTC decoding invents word forms. We therefore decode with a beam search constrained by a lexicon of the 4{,}789 word forms observed in the training transcripts, together with a word 4-gram with stupid backoff estimated on the same transcripts. Candidates are scored by a weighted sum of the CTC and 4-gram log-probabilities and a word-count bonus that offsets CTC's bias towards short outputs, and the $n$-best list is then rescored by a neural language model. 
\section{EXPERIMENTAL SETUP}
\label{sec:setup}

For comparability with the SoniSpeech baseline~\cite{sonispeech}, all models are by default trained on the first 140 sessions of the training split (10.6\,h silent, 10.2\,h voiced). The recognizer is first pretrained on synthetic data alone with the CTC loss for 6 epochs, then trained for a further 60 epochs on real echo profiles together with the units $u$ extracted from the voiced recordings.

\textbf{Generator.} Both generators use AdamW (learning rate $2{\times}10^{-4}$, no weight decay) and are trained for 12 epochs. Stage~1 takes $\lambda_{\mathrm{v}}{=}0.040$, $\lambda_{\mathrm{s}}{=}0.20$ and $\lambda_{\text{adv}}{=}3.0$ in Eq.~\eqref{eq:lg}, and Stage~2 takes $\lambda_{\text{env}}{=}1.0$, $\lambda'_{\text{perc}}{=}0.20$ and $\lambda'_{\text{adv}}{=}3.0$ in Eq.~\eqref{eq:lg2}.

\textbf{Recognizer.} Training uses AdamW (learning rate $3{\times}10^{-4}$, weight decay 0.01, cosine schedule). Group normalization, $16\times$ temporal downsampling, the 1000-piece SentencePiece vocabulary~\cite{sentencepiece} and the SpecAugment and utterance-concatenation augmentations are kept from the baseline~\cite{sonispeech}. CTC targets are lowercased and stripped of punctuation, reproducing the corpus statistics of~\cite{sonispeech} exactly (1{,}684 test word types, 242 out of vocabulary).

\textbf{Evaluation.} We report word error rate (\wer{}) on the full 1{,}000-sentence silent test set, scored after the Whisper English normalizer~\cite{whisper}.
\section{Results}
\label{sec:results}

\begin{table}[t]
\centering
\setlength{\tabcolsep}{4pt}
\caption{Silent-speech \wer{} on the SoniSpeech test set. Rows are cumulative; each addition is applied on top of the row above.}
\label{tab:main}
\begin{tabular}{@{}lccc@{}}
\toprule
Method & Real (sess.) & Syn. (h) & \wer{} (\%) \\
\midrule
SoniSpeech~\cite{sonispeech} & 140 & -- & 26.3 \\
\ours{}                      & 140 & -- & 25.9 \\
\addlinespace[2pt]
\quad + synthetic data       & 140 & 26  & 24.6 \\
\quad + synthetic data       & 140 & 51  & 23.8 \\
\quad + synthetic data       & 140 & 104 & 23.6 \\
\quad + 20 noisy-channel sessions & 160 & 104 & 22.2 \\
\midrule
\quad + lexicon, 4-gram      & 160 & 104 & 18.8 \\
\quad + Qwen3-4B rescoring   & 160 & 104 & 17.1 \\
\quad + 5-model ensemble     & 160 & 104 & \textbf{14.9} \\
\bottomrule
\end{tabular}
\end{table}

\subsection{Overall Performance}
\label{ssec:performance}

Table~\ref{tab:main} builds the system one component at a time. Under the baseline's data budget of 140 sessions, our recognizer alone reads 25.9\% against the published 26.3\%, and pretraining on 104\,h of synthetic profiles lowers this to 23.6\%. The synthetic data is thus worth 2.3 points, over five times the architectural change it sits on.

\textbf{Scale of the synthetic corpus.} We draw 26\,h and 51\,h subsets from the 104\,h corpus, pretrain a recognizer on each and fine-tune on the same 140 real sessions. The gain arrives early and then flattens. A quarter of the corpus already buys more than half of it, 26\,h giving 24.6\%, the doubling to 51\,h a further 0.8 points, and the last doubling to 104\,h only 0.2.

\textbf{Sessions with a faulty microphone.} SoniSpeech~\cite{sonispeech} releases 160 sessions but excludes 20 silent and 11 voiced sessions in which one of the two microphones is dominated by noise. Each microphone contributes two of the four echo-profile channels, one per transmitter, and in the released data the faulty microphone's two channels have been replaced by copies of the intact microphone's, so these sessions carry only two independent channels. Since such a failure is exactly what a device worn daily will meet, we ask whether the data can still be used. 

Starting from the same 104\,h pretrained checkpoint, we fine-tune on all 160 sessions and add an augmentation that overwrites the two channels of one microphone with those of the other, with probability 0.15 per utterance and resampled every epoch. Table~\ref{tab:extradata} separates the two factors. The augmentation alone is worth 0.1 points and does nothing when no degraded data is present, while the 20 extra sessions are worth 0.8; the two together reach 22.2\%, so the augmentation pays off only once the recognizer has degraded sessions to learn from.

\textbf{Decoding.} The remaining errors are largely invented word forms and implausible word sequences, which a lexicon and a language model can repair. Constraining a 128-wide beam search to the training lexicon and adding the word 4-gram removes 3.4 points, and rescoring with Qwen3-4B-Base removes a further 1.7\%. An ensemble of five runs that differ only in random seed brings the system to 14.9\%, a 43\% relative reduction over the baseline. This is, to our knowledge, the first time open-vocabulary silent-speech recognition from an eyewear-mounted sensor has fallen below 15\% \wer{}, the range within which users' willingness to adopt silent speech input holds at the level of an error-free system~\cite{pandey2021acceptability}.

\begin{table}[t]
\centering
\setlength{\tabcolsep}{6pt}
\caption{Contribution of the 20 sessions with a faulty microphone and of the channel-duplication augmentation. All rows start from the same 104\,h pretrained checkpoint.}
\label{tab:extradata}
\begin{tabular}{@{}lcc@{}}
\toprule
& \multicolumn{2}{c}{\wer{} (\%)} \\
\cmidrule(l){2-3}
Training sessions & w/o dup. & w/ dup. \\
\midrule
140 (intact only)      & 23.6 & 23.5 \\
160 (incl. 20 faulty)  & 22.8 & \textbf{22.2} \\
\bottomrule
\end{tabular}
\end{table}

% \begin{table}[t]
% \centering
% \setlength{\tabcolsep}{6pt}
% \caption{Effect of the crossCon objective on the recognizer, without pretraining and on 140 real sessions.}
% \label{tab:crosscon}
% \begin{tabular}{@{}lcc@{}}
% \toprule
% Method & \wer{} (\%) & $\Delta$ (\%) \\
% \midrule
% CTC $+$ crossCon & \textbf{25.9} & -- \\
% CTC only         & 29.0 & $+3.1$ \\
% \bottomrule
% \end{tabular}
% \end{table}

\subsection{Ablation Studies}
\label{ssec:ablation}

\ours{} supervises the recognizer with contrastive learning and the generator with several objectives at once. We remove them one at a time to attribute the gain.

\textbf{Recognizer.} Removing crossCon, and with it the latent-space alignment between audio and echo profiles, raises \wer{} from 25.9\% to 29.0\% on 140 sessions without pretraining. The two modalities can thus be bridged without warping either the audio or the silent profiles explicitly.

\textbf{Generator.} Signal generators are commonly evaluated by the \wer{} of their own output and by similarity to real signals~\cite{stegan}, but neither settles what a generator contributes as supervision. We therefore report both, the downstream \wer{} of a recognizer pretrained on 104\,h of each variant's data, and the \wer{} at which a frozen ResNet-34 judge reads 200 generated silent profiles from the SoniSpeech test set or from the subtitle corpus. Since the generator is supervised through a frozen recognizer, the judge is deliberately a different architecture, trained independently on the same 140 sessions and without the crossCon objective, so that the perceptual loss cannot satisfy it by a shortcut.

Pretraining of any kind adds tens of thousands of optimizer steps, which alone could account for the gain. Two controls separate this from the data itself. The first pretrains on noise matched to the per-channel, per-range-bin mean and variance of real profiles, which reproduces their amplitude statistics while encoding nothing about time. The second replays the real training profiles through the synthetic branch, adding passes over data the recognizer already sees without adding new sentences. Both are run for the same number of optimizer steps and fine-tuned on the same 140 sessions. Neither helps; both hurt, noise by 1.7 points against no pretraining at all and replayed profiles by 0.3. The gain is therefore neither the extra steps nor more passes over the same recordings (Table~\ref{tab:controls}).

On the generation side, the generated profiles are as readable as the real ones. The judge reads generated silent profiles of the SoniSpeech test sentences at 29.0\% against 30.4\% for the real profiles of the same sentences, and every ablation except one stays within 1.1 points of that. Removing $\mathcal{L}_{\text{perc}}$ is the exception and collapses the output to 96.2\%, leaving profiles that are articulation-shaped but unreadable. On the subtitle corpus the same profiles are read 23 points worse. The judge is itself trained on SoniSpeech transcripts, so subtitle-style sentences are out of domain for it as well, and this gap bounds rather than measures the generator's degradation.

Removing each supervision in turn exposes a mismatch between how readable the synthetic data is and how useful it is downstream. Every term contributes downstream, each worth between 0.4 and 0.9 points, yet data whose semantic content has collapsed still improves the recognizer by 1.8 points over no pretraining. Synthetic supervision therefore helps along two routes at once, warming the recognizer up on synthesized articulatory dynamics and carrying the content those dynamics encode, and only the second of the two is visible to a readability metric.

\begin{table}[t]
\centering
\setlength{\tabcolsep}{4pt}
\caption{Ablations of the pretraining data and of the generator's supervision.}
\label{tab:controls}
\begin{tabular}{@{}lccc@{}}
\toprule
& \multicolumn{2}{c}{Synthetic \wer{} (\%)} & Downstream \\
\cmidrule(lr){2-3}
Pretraining data & SoniSpeech~\cite{sonispeech} & Corpus~\cite{danescu2011chameleons, lison2016opensubtitles2016} & \wer{} (\%) \\
\midrule
None                             & --   & --   & 25.9 \\
Noise                            & --   & --   & 27.6 \\
Real data                        & 30.4 & --   & 26.2 \\
\midrule
\ours{}                          & 29.0 & 52.0 & \textbf{23.6} \\
w/o $\mathcal{L}_{\text{env}}$   & 28.3 & \textbf{50.8} & 24.3 \\
w/o $\mathcal{L}_{\text{perc}}$  & 96.2 & 96.6 & 24.1 \\
w/o $D^{\mathrm{p}}_{m}$         & 29.9 & 53.0 & 24.0 \\
w/o $D^{\mathrm{q}}_{m}$         & \textbf{27.8} & 53.2 & 24.5 \\
\bottomrule
\end{tabular}
\end{table}

\section{CONCLUSION}
\label{sec:conclusion}

We presented \ours{}, which synthesizes ultrasonic echo profiles from text and pretrains a silent-speech recognizer on them. The generated profiles are as readable as real recordings, and pretraining on 104\,h of them is worth 2.3 points. Ablations show that readability and downstream usefulness are only loosely coupled, which also explains why the gain saturates early. With the discarded faulty sessions, contrastive learning and standard decoding, the system reaches 14.9\% \wer{}.

\bibliographystyle{IEEEbib}
\bibliography{refs}

\end{document}